# Incoherent lateral shearing digital holographic microscopy

Jaromír Běhal[1,*] and Miroslav Ježek[1]

[1]*Department of Optics, Faculty of Science, Palacký University, 17. listopadu 12, 77146 Olomouc, Czechia*
*jaromir.behal@upol.cz

**Abstract:**
The ability to resolve and quantify features at submicrometer scales from a single-shot image is crucial for real-time uncovering intricate structures in unlabeled biological samples and analyzing them at the subcellular level. We introduce a novel incoherent quantitative phase imaging technique based on lateral shearing digital holographic microscopy, with shearing distances exceeding the coherence area of the employed illumination – a regime not previously achieved and exploited. This strategy consequently assures artifacts-free, high-accuracy, and high-resolution quantitative phase reconstructions. This regime extends the single-shot lateral shearing digital holographic microscopy towards incoherent illumination, thus increasing the space-time bandwidth product of the method. The practical common-path geometry also provides enhanced vibration resistance, which is essential for consistent time-lapse measurements. We verify the scalability and effectiveness of the method by investigating samples via multiple condenser-objective pairs, providing a wide range of lateral resolutions and fields of view, thereby assuring its applicability for various microscope settings. We experimentally verify long-term phase stability and unprecedented phase-reconstruction accuracy. Finally, we use our method for investigation of biological samples, including cheek cells, diatoms, and yeast cells, highlighting its potential for dynamic label-free analysis at the organelle level.

## 1. Introduction

Digital holographic microscopy (DHM) is a versatile coherence-based method that delivers detailed quantitative phase information from various sample depths at the microscopic or even nanoscopic levels. DHM has become a state-of-the-art technique that enables precise, non-contact, non-destructive sample analysis with applications spanning various research fields, including biomedicine and material sciences [1–5]. It also excels in capturing data under extreme and complex conditions, such as underwater, turbid, or scattering/diffuse environments [6–9]. Common DHM strategies utilize interference patterns formed by a mixture of the mutually coherent object and reference optical fields. The stack of recorded patterns is further processed, and the retrieved complex amplitude reveals the optical path difference induced by the investigated object.

Most DHM approaches rely on temporally and spatially coherent illumination, which brings severe issues, including increased speckle noise, parasitic interference artifacts, and reduced resolution of the imaging systems [10]. Indeed, only a few single-shot incoherent holographic methods enable true full-field-of-view quantitative phase analysis. The full spatiotemporal overlap between two off-axis signal and reference waves originating from one partially coherent light source can be achieved when incorporating a diffraction grating, a prism, or appropriate volume diffractive optical element into the reference arm of the interferometer [11–13]. This strategy keeps the coherence planes of both mutually tilted waves parallel, hence the uniform interference is preserved across the whole field of view (FoV). The above techniques extend applications of the single-shot off-axis DHM towards incoherent illumination, with a consequent boost of the lateral resolution and homogeneity of the reconstructed quantitative phase map.

Common two-arm interferometers, such as the Mach-Zehnder configuration, provide sample-independent reference waves at the cost of increased setup complexity, reduced phase stability without external stabilization, and lower imaging accuracy in particular interferometric designs [14]. These limitations may hinder practicality of DHM for required real-word applications. Enhanced phase stability can be achieved using common-path off-axis DHM. From the perspective of the approach to generate the reference beam, we categorize the common-path models mainly as point diffraction DHM and lateral shearing DHM (LSDHM), respectively [15]. The point-diffraction arrangements utilize interference between two fully overlapping copies of the FoV. A limitation associated with the sample-dependent reference wave is addressed via tight spectral filtering on one of the FoV copies, which is usually achieved through the precisely positioned pinhole. The filtering process permits the transition of low spatial frequencies of the sample, hence creating a 'clean' reference wave that interferes with the (second) unaffected copy of the FoV. Such holographic approach allows imaging of low-absorbing samples. In particular, the point diffraction DHM, developed in work [16], used a diffraction grating to assure spatiotemporal overlap between the interfering waves in addition to the FoV splitting. Such a method, called diffraction phase microscopy, allowed experiments under spatially incoherent and white-light illumination [17, 18]. However, if the spatial filtering is inadequate, non-zero spatial frequencies leak into the reference field, so the final holographic reconstruction lacks the corresponding non-zero spectral content. Consequently, the shaded-off and halo artifacts appear, underestimating the reconstructed phase [19].

On the other hand, the simpler-to-adjust lateral-shearing approaches exploit self-interference between two mutually displaced (sheared) copies of the FoV [20]. The technique can be used for the phase-gradient retrieval [21, 22], which provides wavefront derivative along the shear direction, or for the direct DHM reconstruction [23]. This manuscript focuses on the direct DHM, extracting the quantitative phase distribution across the entire object rather than the phase-gradient retrieval. Indeed, the direct holographic imaging is accomplished by increasing the shear beyond the dimensions of an observed object, leading to a straightforward DHM. Especially in LSDHM, the empty part of the first FoV replica serves as a reference wave for the second replica, which is advantageous for sparse and/or bounded samples, a typical scenario in microfluidics [24]. LSDHM thus provides quantitative phase reconstructions without shaded-off and halo artifacts. Nevertheless, the finite lateral shear distance in such a cross-referenced interference strategy restricts the applications to the fully or partially coherent illumination with a coherence area larger than the shear distance. According to the authors' knowledge, no widely recognized study explicitly demonstrates a successful incoherent LSDHM technique where the shear distance significantly exceeds the coherence area; hence, the use of spatially incoherent illumination in common-path LSDHM has yet to be introduced.

In this work, we present a novel incoherent quantitative phase imaging technique based on LSDHM with the shearing distances exceeding the coherence area of the employed illumination. This setting provides artifacts-free, high-accuracy, and high-resolution quantitative phase reconstructions. Significantly, this approach simplifies the implementation of incoherent digital holographic microscopy by eliminating the need for two-arm interferometers, thereby offering an efficient and easily reconfigurable solution for quantitative-phase imaging systems. The practical common-path geometry ensures advanced vibration resistance, a crucial feature for consistent time-lapse measurements. We highlight the adaptability of the holographic microscope by using various high-numerical-aperture condenser-objective pairs, providing various FoVs and optical resolutions. Furthermore, we experimentally verify long-term phase stability and unprecedented phase-reconstruction accuracy. Additionally, we image single cells in high resolution, thus providing single-shot label-free analysis at the organelle level.

## 2. Core idea

Fig. 1 illustrates relations between the sample and image planes for major common-path off-axis DHM geometries. In the configuration typical for the point diffraction DHM, especially diffraction phase microscopy (Fig. 1(a)), the sample plane is duplicated by a configuration comprising a diffraction grating followed by a 4f optical system with a built-in spatial filter. Here, one copy remains unaffected, serving as an object wave, while the second one is low pass filtered, acting as a reference wave. This configuration allows experiments with spatially incoherent and broadband light sources as the coherence planes of both waves remain perpendicular to the optical axis (illustrated as gray rectangles in Fig. 1(a)). The waves consequently interfere creating interference fringes across whole FoV, but only if the signal wave 100% spatially overlaps its low-pass filtered replica, i.e., situation when the lateral shear distance ($s'$) is zero ($s' = 0$).

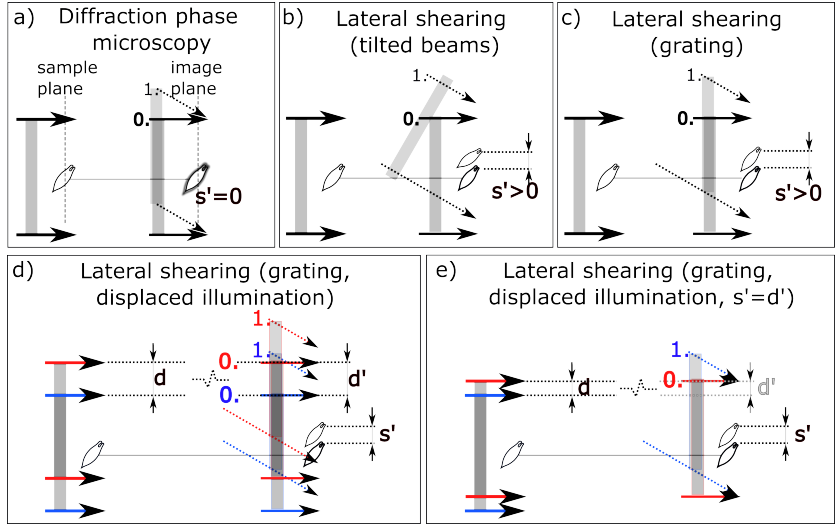

Fig. 1. Sample plane (left) and image plane (right) in common-path off-axis DHM configurations. Arrows indicate light-propagation direction while grey rectangles denote tilt of the coherence plane. a) Point diffraction DHM, especially diffraction phase microscopy. b) LSDHM with tilted beams. c) LSDHM with diffraction grating. d) LSDHM from (c) with two parallel mutually displaced illumination beams ($d' \neq s'$). e) LSDHM from (d) with when $d' = s'$. Symbols d,d' are the displacement distances in the object and the image space, respectively, while $s'$ is a lateral shear distance in the image space.

Moreover, in a typical LSDHM configuration (Fig. 1(b)), one copy of the FoV remains unaffected, serving as an object wave, while the second copy is displaced and tilted without any spatial filtration. Hence, the sample-free area of the first copy acts as a reference wave for the second copy and vice versa. The spatial point uncorrelation between both replicas ($s' > 0$) does not allow to use the method for spatially incoherent illumination, as significant lateral shear distances $s'$ must be introduced to avoid the overlap between the two sample images. Furthermore, a mismatch between the coherence-plane tilts restricts the use of the method for illumination with a narrow spectrum, which assures sufficient temporal coherence necessary for contrast interference fringes across whole FoV.

A similar situation appears in Fig. 1(c), where the lateral shear distance $s'$ is induced by axial displacement of the diffraction grating placed before a 4f optical system as proposed in the reference [25]. The distance $s'$ set by this shearing module is proportional to the axial displacement of the grating from the intermediate image plane (primary image plane) of the microscope. The experimental details are not shown in Fig. 1(c) but they will be introduced and discussed later in detail. The spatial point uncorrelation between the replicas in Fig. 1(c) also does not allow to use this technique for spatially incoherent illumination. Nevertheless, it is worth pointing out that the coherence planes remain parallel compared to the situation in Fig. 1(b).

Furthermore, let us consider Fig. 1(c) with two parallel equal-wavelength illumination beams displaced of the distance d in the sample plane. The beams are labeled red and blue for illustration purposes in Fig. 1(d). The output (red and blue) optical fields are displaced correspondingly of the distance d′, while the lateral shear distance s′ is generally different (d′ ≠ s′) and independent of d. Especially, changes of the illumination displacement d do not influence the lateral shear distance s′. An interesting situation occurs when d′ = s′. In this case, the red zeroth diffraction order spatiotemporally overlaps the first diffraction order of the blue illumination beam. Such a prediction is present in Fig. 1(e), where partially overlapping optical fields have been filtered out for convenience. Nevertheless, the interference cannot appear unless both light beams originate from the same incoherent light source.

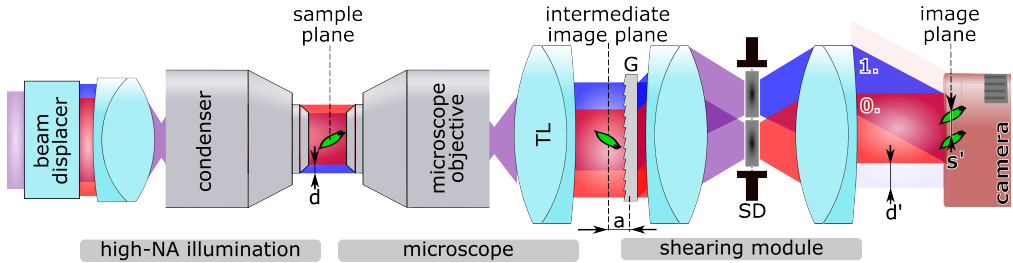

Fig. 2. Simplified experimental layout consisting of a beam displacer, a high-NA illumination, a microscope, a shearing module and a camera. Tube lens (TL), diffraction grating (G), separation diaphragm (SD), illumination displacement in the sample plane (d) and the image plane (d′), lateral shear distance in the image plane (s′), axial displacement of the grating (a).

Fig. 2 proposes a simplified layout satisfying all the above requirements with all the lenses arranged in a telecentric configuration. The initial incoherent illumination is duplicated and displaced without tilt by a beam displacer. The subsequent high-numerical-aperture (high-NA) sample illumination, consisting of a lens and a high-NA condenser, reimages the displaced incoherent light field into the sample plane. The first sample image arises in the intermediate image plane of a microscope, consisting of a microscope objective and a tube lens (TL). The following shearing module provides an adjustable lateral shear distance s′ in a camera plane (image plane), which comprises the diffraction grating (G) and a 4f optical system. The s′ is set via axial shift (a) of the G away from the intermediate image plane. The separation diaphragm (SD) transmits the zeroth and the first diffraction orders of G while the remaining diffraction orders remain blocked. The inserted filters inside SD filter out the unnecessary light fields from the zeroth and the first diffraction orders, respectively, to reduce an unwanted intensity bias in the image plane (Fig. 1(e), Fig. 2), and the final interference pattern is recorded by the camera.

## 3. Results

The realization and detailed description of the experimental configuration proposed in Fig. 2 can be found in the Methods. Fig. 3 provides experimental results confirming the working principle of the method. Here, the illumination has been realized by an incoherent fluorescence-based light source coupled into the light guide, which is followed by two additional lenses. The output light field is subsequently displaced via an adjustable beam displacer and re-imaged into the sample plane. The first used condenser-objective arrangement comprises a 20×/0.4 condenser and a 10×/0.25 microscope objective. In Fig. 3(a), the direct image of a negative high-resolution USAF 1951 resolution target is present, which was recorded when only the zeroth diffraction order of G passed towards the camera, while the first diffraction order was blocked. Cross-section through the finest resolved lines of the image, plotted in Fig. 3(b), demonstrates decaying visibility with increasing spatial frequency of the lines. It can be observed that the spatial frequency corresponding to the first unresolved lines (element 6, group 9 = 912 lp/mm) agrees with the theoretical resolution limit given by the largest spatial frequency transmitted via the incoherent optical imaging system, which is given as $2NA/\lambda$ = 940 lp/mm, where NA is numerical aperture of the used microscope objective, thus confirming low-spatial coherence of the used illumination.

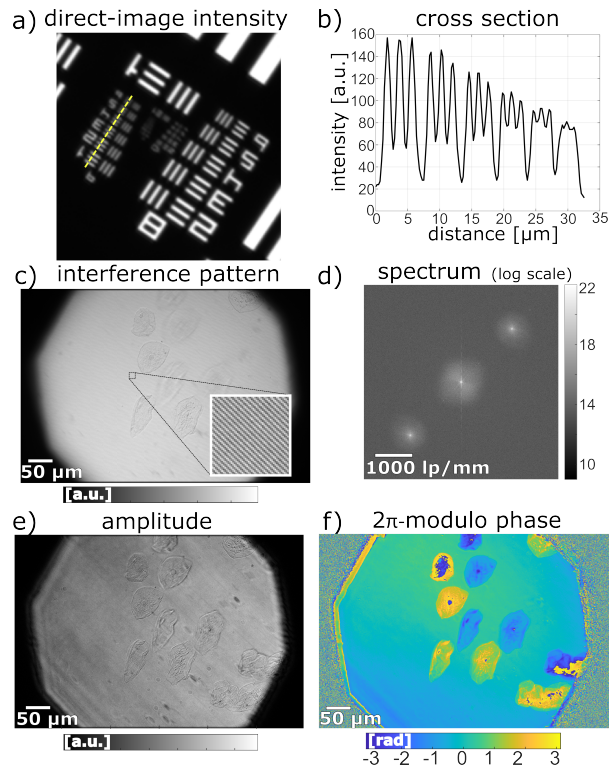

Fig. 3. Results for a 20×/0.4 condenser and a 20×/0.25 microscope objective. a) Direct image of line target USAF 1951. b) Cross-section through the smallest resolved lines (group 9). c) Interference pattern with cheek cells as a sample. d) Absolute value of Fourier spectrum in logarithmic scale. e) Reconstructed amplitude. f) Reconstructed $2\pi$-modulo phase map with clear complex-conjugated replicas.

The holographic performance was tested on cheek cells scratched from the inside of a mouth by a toothpick, diluted in a drop of water and sandwiched between the glass slide and the cover glass. The interference pattern produced by both transmitted diffraction orders is present in Fig. 3(c) with the inset image showing emerged interference fringes. Fig. 3(d) shows the Fourier spectrum of the interference record. The reconstructed amplitude is shown in Fig. 3(e) together with the corresponding $2\pi$-modulo phase map in Fig. 3(f), both with apparent complex-conjugated twin images. The lateral shear distance in the sample plane for the proposed setting reaches $s \approx 100$ $\mu$m. Furthermore, background phase distortions consisting mainly of residual tilt and quadratic phase (not shown in the figures) were corrected by the reference interference pattern recorded in the sample-free area before measurements with cells [26].

The variability and scalability of the method were verified by investigating samples in multiple condenser-objective configurations. After the initial configuration, a 20×/0.4 condenser and a

20×/0.4 microscope objective (Fig. 4) were used to observe a diatom from the micromanipulated diatom slide (XL Diatomarium, Diatom shop). Square root of directly captured intensity when blocking the first diffraction order in the SD-diaphragm plane (Fig. 4(a)), is compared to the reconstructed amplitude (Fig. 4(b)). The reconstructed phase (Fig. 4(c)) contains a homogenous background and the diatom with a comb-like structure consisting of approximately 1$\mu$m distant spikes. Both attributes are evident in the cross-section plot (Fig. 4(d)) along the red-colour line in Fig. 4(c). Importantly, the results are obtained without any phase-smoothing, thresholding, or denoising.

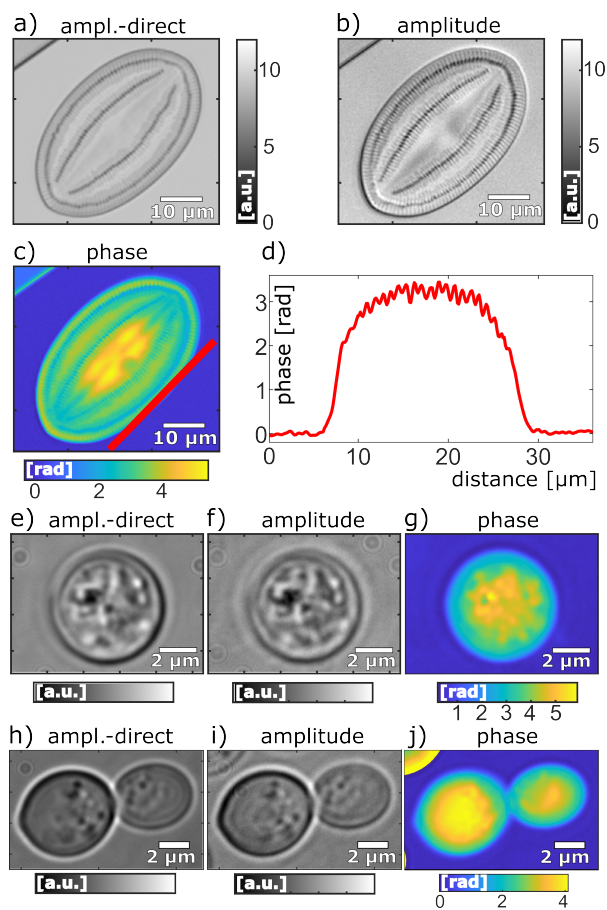

Fig. 4. Diatom: Results for a 20×/0.4 condenser and a 20×/0.4 microscope objective. a) Square root of recorded intensity b) Reconstructed amplitude. c) Reconstructed phase. d) Phase-profile cross-section. Yeast cells: Results for a 100×/0.90 condenser and a 100×/1.40 oil-immersion microscope objective. e) Square root of recorded intensity. f) Reconstructed amplitude. g) Reconstructed phase. Results for a 100×/1.40 oil-immersion condenser and a 100×/1.40 oil-immersion microscope objective. h) Square root of recorded intensity. i) Reconstructed amplitude. j) Reconstructed phase.

Moreover, a 100×/0.9 condenser – 100×/1.4 oil–immersion microscope objective (Fig. 4(e), (f), (g)) and a 100×/1.4 oil-immersion condenser – 100×/1.4 oil-immersion microscope objective (Fig. 4(h), (i), (j)) configurations were used for detailed observation of yeast cells, which were obtained by mixing compressed fresh baker's yeast with room-temperature water and sugar. A drop of such mixture was sandwiched between two cover glasses and used as a sample (used immersion oil – Cargille laboratories, type FF). In both cases, the square root of the captured intensities (Fig. 4(e), (h)) is compared to the reconstructed amplitudes (Fig. 4(f), (i)) and the corresponding phase maps (Fig. 4(g), (j)), concluding that fine structures observed in the direct images are preserved in the reconstructed amplitudes along with the quantitative phase maps of a single cell (Fig. 4(g)) and two dividing cells (Fig. 4(j)), respectively.

## 4. Methods

### 4.1. Detailed setup description

The experimental layout is comprehensive, comprising several function blocks, i.e., a low-numerical-aperture (low-NA) Köhler illumination, a Sagnac displacer, a high-NA Köhler illumination, a microscope, and a shearing module (Fig. 5(a)). The initial Köhler illumination path exploits a broadband incoherent light produced by a fluorescent light source (LS; Crytur MonaLIGHT B01, central wavelength 555 nm) coupled into a 4 mm diameter light guide, diaphragms, and achromatic doublets in telescope configuration. In particular, the low-NA Köhler illumination contains lenses $L_1$ ($f_1$=30 mm), $L_2$ ($f_2$=50 mm), and a condenser lens $C_1$ ($f_{C1}$=100 mm). The field diaphragm (FD) is placed in the common focal plane of lenses $L_1$ and $L_2$, while the aperture diaphragm (AD) is placed in the common focal plane of lenses $L_2$ and $C_1$. The subsequent lens $L_3$ ($f_3$=150 mm) and a high-NA condenser $C_2$ (Olympus 20×/0.40, or Olympus 100×/0.90, or Olympus 100×/1.40 oil-immersion) reimage the low-NA illumination into the sample plane (S), thus achieving the high-NA illumination. The adjustable beam displacer, inserted between $C_1$ and $L_3$, is realized by a triangular Sagnac geometry, which consists of a polarizing beam splitter (PBS), a fixed mirror ($M_1$), and a movable mirror ($M_2$) placed on a linear translation stage. The mirror translation determines the displacement d between the arisen high-NA illumination copies in the sample plane, while the linear polarizer $LP_1$ defines the input polarization state, thus balancing the intensity ratio between the passing-through beams.

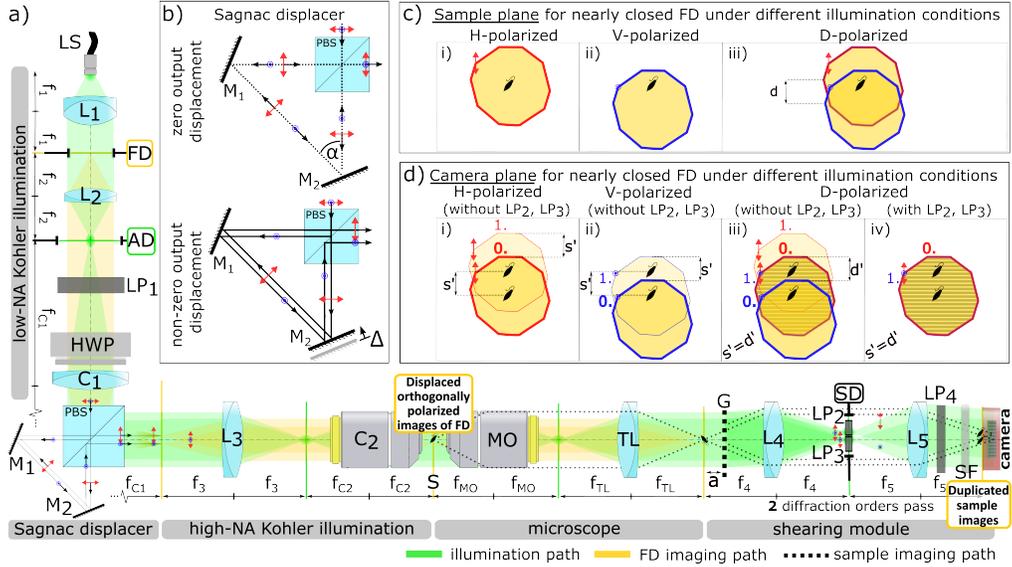

Fig. 5. a) Sketch of the experimental layout. b) Working principle of triangular Sagnac displacer. c) Sample plane under different settings with reimaged nearly closed FD. d) Camera plane under different settings. FD imaging path is omitted in the shearing module geometry for the illustration simplicity.

The magnified sample image arises in the intermediate image plane of a microscope, consisting of a microscope objective MO (Olympus 10×/0.25, or Olympus 20×/0.40, or Olympus 100×/1.40 oil-immersion) and a tube lens TL ($f_{TL}$=300 mm). The following shearing module provides an adjustable lateral shear distance s′ in a camera plane by translating the grating from the intermediate image plane of distance a. The shearing module consists of the diffraction grating G (Ronchi 120 lp/mm, or Ronchi 80 lp/mm, respectively) and a telescope with lenses $L_4$ ($f_4$=200 mm) and $L_5$ ($f_5$=150 mm). The SD placed in the back focal plane of $L_4$ fully transmits the zeroth and the first diffraction orders of the grating G while blocking all remaining diffraction orders. The linear polarizers $LP_2$ and $LP_3$ inserted inside SD transmit horizontally-polarized (H-polarized) and vertically-polarized (V-polarized) light contributions, respectively, hence filter out unnecessary polarization components present in the zeroth and first diffraction orders, thus lowering an unwanted intensity bias in the camera plane (Fig. 1(e), Fig. 5(div)). It is worth noting that no precise pinhole positioning is needed for the spectral filtering purposes. The linear polarizer $LP_4$ oriented diagonally assures interference of the both overlapping orthogonally polarized light fields. The camera (ZWO-ASI 178MM, 2.4 $\mu$m square pixels) with a mounted spectral filter (central wavelength $\lambda$=532 nm, full width at half maximum FWHM = 4 nm) captures the final interference pattern.

The working principle of the method is depicted in Fig. 5(c),(d), where nearly closed FD under different illumination conditions is present with an emphasis on the two displaced orthogonally polarized replicas of the initial illumination. The situation for H-polarized sample illumination is present in Fig. 5(ci), which shows the FD and a small black sample. The subsequent optical path, comprising the microscope and the shearing module (Fig. 5(a)), shears and images the observed FoV into the camera plane (Fig. 5(di)). The shear direction follows the G orientation, and the shear distance in the sample plane s depends on the grating period $g_P$ according to equation [25]

$$s = \frac{s'}{\beta} = \frac{a\lambda}{g_P} \frac{f_{MO}}{f_{TL}}, \tag{1}$$

where s′ is lateral shear distance in the camera plane, $\lambda$ is the central wavelength of the transmitted light, and $\beta = (f_{TL} f_5)/(f_{MO} f_4)$ is a lateral magnification of the whole sample-imaging optical path. The sheared images originate from the zeroth and the first diffraction order of the G; thus, the setup works as a LSDHM. Nevertheless, this configuration cannot provide any interference pattern for spatially incoherent illumination if s′ ≥0 (Fig. 1(c)). A similar situation arises for V-polarized illumination (Fig. 5(cii)) with the corresponding camera-plane image (Fig. 5(dii)). The sample and the sample-image positions remain, but the FD images are shifted compared to the previously discussed H-polarized case. This shift is represented by the distance d, as evident in Fig. 5(ciii), which shows the situation when the diagonally polarized (D-polarized) beam enters the Sagnac displacer (Fig. 5(b)). In this scenario, both H-polarized and V-polarized components illuminate the sample with the displacement d given by the formula

$$d = \frac{d'}{\beta} = \frac{f_{C2}}{f_3} \frac{2\Delta \sin(\alpha)}{\sin(\pi/2 - \alpha/2)} \approx \frac{f_{C2}}{f_3} 1.53\Delta \text{ for } \alpha = \pi/4, \tag{2}$$

where d′ is displacement in the camera plane, $\alpha$ is the angle in the triangular Sagnac geometry, and $\Delta$ is displacement of the mirror $M_2$ from the zero-output displacement position (Fig. 5(b)). Importantly, the lateral shear distance s and the illumination displacement d are independent. If the orientation and size of the lateral shear s and the displacement d are equal, the zeroth order of one polarization component spatially overlaps the first order of the orthogonally polarized component in the camera plane. This situation (s′ = d′) is depicted in Fig. 5(diii), where the H-polarized zeroth diffraction order overlaps the V-polarized first diffraction order. In such a case, the interference pattern arises inside the 100%-overlap area in the camera plane because the linear polarizer $LP_4$ projects H and V components into the common D-polarization direction. Furthermore, Fig. 5(diii) illustrates that both diffraction orders related to both polarization components are transmitted, i.e., four light fields in total. The additional linear polarizers LP$_2$ and LP$_3$ inserted inside the SD, filter out the non-interfering parts, causing the intensity bias. Notably, the H-polarized component from the first and the V-polarized component from the zeroth diffraction orders are blocked, while the remaining polarization components pass (note also Fig. 1(e)). The final camera record in such a situation is sketched in Fig. 5(div).

It is important to note that the interference appears only if the optical path difference (OPD) between the interfering waves [27]

$$\text{OPD} = a\left(\frac{1}{\cos(\lambda/g_P)} - 1\right) \quad (3)$$

does not exceed the illumination coherence length for Gaussian spectrum given as [28]

$$l_{coh} = \sqrt{\frac{2\ln(2)}{\pi}}\frac{\lambda^2}{\text{FWHM}}, \quad (4)$$

where FWHM is the full width at half maximum of the spectrum, or the OPD is appropriately compensated, e.g., by the external cover glass slide placed into the first diffraction order in the SD-diaphragm plane.

### 4.2. Reference measurements

Phase accuracy was established for a 100×/1.4 oil-immersion condenser and a 100×/1.4 oil-immersion microscope objective using glass beads (10 – 30 $\mu$m diameter, Polyscience, # 07688; refractive index at 532 nm: $n_{beads} = 1.5168$ [29]) dispersed in oil (Cargille laboratories, type FF, refractive index at 532 nm: $n_{oil} = 1.4819$) as a sample. The reconstructed complex amplitudes were numerically refocused to the optimal sample plane, determined as the position with the lowest amplitude contrast. The contrast was evaluated in the region containing the border between the glass beads and the surroundings [30]. The reconstructed $2\pi$-modulo phase of a representative bead is shown in Fig. 6(a), while the unwrapped phase is presented in Fig. 6(b). The whole two-dimensional unwrapped phase was further fitted by a phase difference induced by the ideal sphere in the form

$$\delta\phi = \frac{4\pi\delta n}{\lambda}\Re\left[\sqrt{(D/2)^2 - (x - x_0)^2 - (y - y_0)^2}\right] + \delta\phi_0, \quad (5)$$

where $\delta n$ is the difference between refractive index of the glass bead and the surrounding medium, $D$ is a diameter of the sphere, $\{x_0, y_0\}$ is coordinate of its centre, $\delta\phi_0$ is a possible phase-background offset, and $\Re[]$ represents real part of the complex number. In particular, the nonlinear data-fitting problem was solved using the function lsqcurvefit of the software Matlab (R2024a). The cross-section through the fitted phase profile (blue dashed line in Fig. 6(c)) is compared with the two cross-sections through the unwrapped phase (red and green lines in Fig. 6(c)), verifying high accuracy of the reconstructed phase. Indeed, the evaluated R-square parameter ($R^2 = 99.94$ %) confirms the outstanding goodness of the fitted model. Overall, five different glass beads with diameters ranging from 14.1 $\mu$m to 20.5 $\mu$m were examined, yielding a final goodness $R^2 = 99.92 \pm 0.04$ %. Additionally, the estimated refractive index difference between the beads and the surrounding oil ($\delta n = 0.0367 \pm 0.0006$) agrees with the expected value ($n_{beads} - n_{oil} = 0.0349$), although the evaluation precision may be affected by imperfections arising from variations in the material properties or geometry.

Moreover, the temporal stability was evaluated in the same experimental layout by recording a stack of 200 interference patterns during 12 minutes with 2-second exposure. The measurements were accomplished following the real experimental conditions, i.e., with two attached cover glasses in the object space together with the double oil immersion. Subsequently, phase maps were reconstructed from all images in the stack. The standard deviation distribution evaluated in the area 19.2×19.2 $\mu m^2$ is present in Fig. 6(d). A mean standard deviation of 0.023 rad corresponds to a 1.9nm OPD, which confirms the high temporal stability of the common-path interferometer even for the double oil-immersion configuration without any vibration compensation.

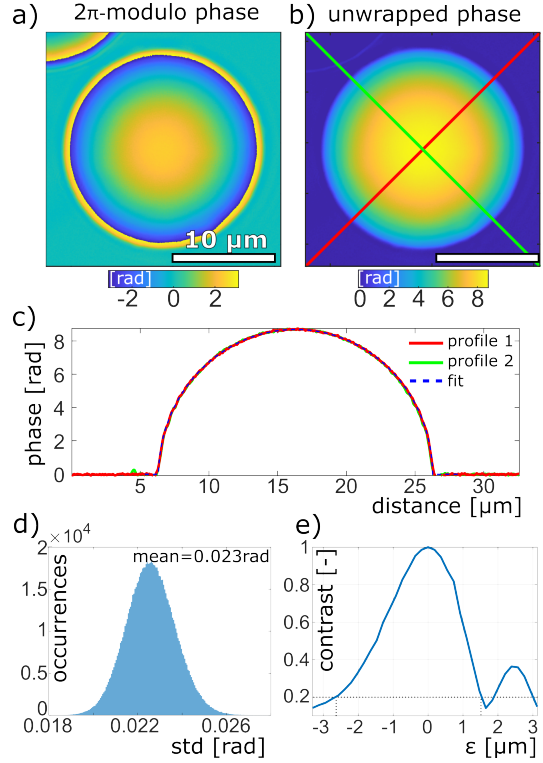

Fig. 6. Results for a 100×/1.40 oil-immersion condenser and a 100×/1.4 oil-immersion microscope objective. Glass beads diluted in immersion oil: a) Reconstructed $2\pi$-modulo phase. b) Unwrapped phase. c) Phase profile through the red and green lines marked in b) together with fitted geometric phase profile induced by the ideal sphere (blue dashed line). d) Temporal stability. Mean standard deviation of the reconstructed phase in the area 1000×1000 pixels achieves 0.023 rad, corresponding to 1.9nm OPD in the area 19.2×19.2 $\mu m^2$ = 369$\mu m^2$. e) Interference-fringe contrast as a function of the displacement $\epsilon$ from the position with the maximal fringe visibility d = $d_0$.

Finally, we estimated spatial coherence area by displacing orthogonally polarized illumination replicas behind the Sagnac interferometer (Fig. 5(b)). The illumination displacement in the sample plane d was set by moving the mirror $M_2$ of distance $\Delta$ (Eq. 2). Fig. 6(d) shows the interference-fringe contrast, evaluated in the central 200×200 pixels of the camera, as a function of the displacement $\epsilon$ from the position with the maximal fringe visibility d = $d_0$. Hence, the current illumination displacement in the sample plane can be expressed as d = $d_0$ + $\epsilon$. The graph shows the fringe visibility decreases to approximately 20 % of its maximum for $\epsilon \approx 2$ $\mu$m, which corresponds to the spatial coherence radius. Indeed, the estimated coherence area is an order of magnitude smaller than the lateral shear, which is s = $d_0$ = 22 $\mu$m in the current experimental setting.

## 5. Conclusion

Lateral shearing digital holographic microscopes allow direct quantitative phase imaging by increasing the shear beyond the dimensions of an observed object, leading to a robust common-path imaging system suitable for compact portable devices. However, the imaging quality of these microscopes has been limited by spatially coherent illumination, leading to an uneven speckled phase and a resolution constrained by spatially coherent illumination. Our method overcomes all these limitations and extends the potential of lateral shearing digital holography towards high-accuracy and high-resolution real-world applications. We believe our work will be of broad interest to the scientific community and will open new exciting avenues for practical applications across diverse fields, from material science to biomedicine.

Mainly, this manuscript introduced an incoherent common-path lateral shearing digital holographic microscope that achieves high reconstruction accuracy without phase artifacts. This approach simplifies the implementation of incoherent off-axis digital holographic microscopy by eliminating the need for two-arm interferometers, thereby offering an efficient and easily reconfigurable solution for holographic imaging systems. The working principle involved exploiting the displaced incoherent-illumination replicas in conjunction with the diffraction grating. We demonstrated that full-field-of-view interference appears when the illumination displacement corresponds to the lateral shear distance.

In particular, the incoherent common-path lateral shearing digital holographic microscope was realized for lateral shear distances exceeding the coherence area. We verified the scalability and effectiveness of the method by investigating samples via multiple condenser-objective pairs, providing a wide range of lateral resolutions and fields of view, thereby assuring its applicability for various microscope settings. Notably, it was demonstrated for the double-immersion configuration with two 100×/1.40 oil-immersion microscope objectives that the estimated 2$\mu$m spatial coherence area was ≈ 10× smaller compared to the 22$\mu$m lateral shear distance. Furthermore, the measured phase of five glass beads dispersed in an oil medium was fitted by the sphere-induced phase difference model. Diameters of the beads ranged from 14.1 $\mu$m to 20.5 $\mu$m. The evaluated R-square parameter ($R^2$ = 99.92 ± 0.04 %) confirmed the exceptional goodness of the fit, thus demonstrating the high accuracy of the retrieved phase. Indeed, the estimated refractive index difference between the beads and the surroundings agreed with the expected value. Moreover, temporal stability tests revealed a phase measurement standard deviation of less than 2 nm over 12 minutes without any vibration compensation. Even better results are expected for oil-immersion configurations with horizontally oriented samples, which minimize immersion oil flow during measurements. Additionally, dry condenser-objective combinations paired with microfluidic applications may further enhance phase stability by allowing the reference wave to pass through the liquid-free region of the microfluidic chip. The final measurements with diatoms, human cheek cells, and yeast cells validated the high reconstruction quality of the imaged samples. Notably, the side-by-side comparison of direct light-microscope images of cells with their reconstructed amplitudes proved that fine structures observed in the high-resolution direct images remained in the reconstructed amplitudes along with the accurate quantitative phase maps of the cells.

**Funding.** Grant Agency of the Czech Republic (No. 25-17712I).

**Disclosures.** The authors declare no conflicts of interest.

**Data Availability Statement.** Data underlying the results presented in this paper are not publicly available at this moment but may be obtained from the authors upon reasonable request.

## References

1. Y. Park, C. Depeursinge, G. Popescu, "Quantitative phase imaging in biomedicine," Nat. Photon **12**, 578–589 (2018). https://doi.org/10.1038/s41566-018-0253-x.

2. B. Javidi, A. Carnicer, A. Anand, *et al.*, "Roadmap on digital holography [Invited]," Opt. Express **29**, 35078–35118 (2021). https://doi.org/10.1364/OE.435915.
3. T. L. Nguyen, S. Pradeep, R. L. Judson-Torres, *et al.*, "Quantitative Phase Imaging: Recent Advances and Expanding Potential in Biomedicine," ACS Nano **16**, 11516–11544 (2006). https://doi.org/10.1021/acsnano.1c11507.
4. D. Pirone, J. Lim, F. Merola, *et al.*, "Stain-free identification of cell nuclei using tomographic phase microscopy in flow cytometry," Nat. Photon **16**, 851–859 (2022). https://doi.org/10.1038/s41566-022-01096-7.
5. S. Khadir, D. Andrén, P. C. Chaumet, *et al.*, "Full optical characterization of single nanoparticles using quantitative phase imaging," Optica **7**, 243–248 (2020). https://doi.org/10.1364/OPTICA.381729.
6. S. L. C. Giering, E. L. Cavan, S. L. Basedow, *et al.*, "Sinking Organic Particles in the Ocean—Flux Estimates From in situ Optical Devices," Front. Mar. Sci. **6**, 834 (2020). https://doi.org/10.3389/fmars.2019.00834.
7. A. R. Nayak, M. N. McFarland, J. M. Sullivan, *et al.*, "Evidence for ubiquitous preferential particle orientation in representative oceanic shear flows," Limnol. Oceanogr. **63**, 122–143 (2017). https://doi.org/10.1002/lno.10618.
8. J. Huang, Y. Zhu, Y. Li, *et al.*, "Snapshot Polarization-Sensitive Holography for Detecting Microplastics in Turbid Water," ACS Photonics **10**, 4483–4493 (2023). https://doi.org/10.1021/acsphotonics.3c01350.
9. M. Paturzo, A. Finizio, P. Memmolo, *et al.*, "Microscopy imaging and quantitative phase contrast mapping in turbid microfluidic channels by digital holography," Lab Chip **12**, 3073–3076 (2012). https://doi.org/10.1039/C2LC40114B.
10. V. Bianco, P. Memmolo, M. Leo, *et al.*, "Strategies for reducing speckle noise in digital holography," Light Sci Appl **7**, 48 (2018). https://doi.org/10.1038/s41377-018-0050-9.
11. Y. Choi, T. D. Yang, K. J. Lee, *et al.*, "Full-field and single-shot quantitative phase microscopy using dynamic speckle illumination," Opt. Lett. **36**, 2465–2467 (2011). https://doi.org/10.1364/OL.36.002465.
12. S. Witte S, A. Plauşka, MC. Ridder, *et al.*, "Short-coherence off-axis holographic phase microscopy of live cell dynamics," Biomed Opt Express **3**, 2184–2189 (2012). https://doi.org/10.1364/BOE.3.002184.
13. Z. Monemhaghdoust, F. Montfort, E. Cuche, *et al.*, "Full field vertical scanning in short coherence digital holographic microscope," Opt. Express **21**, 12643–12650 (2013). https://doi.org/10.1364/OE.21.012643.
14. J. Lullin, S. Perrin, M. Baranski, *et al.*, "Impact of mirror spider legs on imaging quality in Mirau micro-interferometry," Opt. Lett **40**, 2209–2212 (2015). https://doi.org/10.1364/OE.21.012643
15. J. Zhang, S. Dai, C. Ma, *et al.*, "A review of common-path off-axis digital holography: towards high stable optical instrument manufacturing," Light Adv. Manuf **2**, 23 (2021). https://doi.org/10.37188/lam.2021.023.
16. G. Popescu, T. Ikeda, R. R. Dasari, *et al.*, "Diffraction phase microscopy for quantifying cell structure and dynamics," Opt. Lett **31**, 775–777 (2006). https://doi.org/10.1364/OL.31.000775.
17. B. Bhaduri, H. Pham, M. Mir, *et al.*, "Diffraction phase microscopy with white light," Opt. Lett. **37**, 1094–1096 (2012). https://doi.org/10.1364/OL.37.001094.
18. B. Bhaduri, C. Edwards, H. Pham, *et al.*, "Diffraction phase microscopy: principles and applications in materials and life sciences," Adv. Opt. Photon. **6**, 57–119 (2014). https://doi.org/10.1364/AOP.6.000057.
19. P. C. Chaumet, P. Bon,G. Maire, *et al.*, "Quantitative phase microscopies: accuracy comparison," Light Sci Appl **13**, 288 (2024). https://doi.org/10.1038/s41377-024-01619-7.
20. L. A. Alemán-Castaneda, B. Piccirillo, E. Santamato, *et al.*, "Shearing interferometry via geometric phase," Optica **6**, 396–399 (2019). https://doi.org/10.1364/OPTICA.6.000396.
21. P. Bon, N. Bourg, S. Lécart, *et al.*, "Three-dimensional nanometre localization of nanoparticles to enhance super-resolution microscopy," Nat Commun **6**, 7764 (2015). https://doi.org/10.1038/ncomms8764.
22. Q. Song, S. Khadir, S. Vézian, *et al.*, "Printing polarization and phase at the optical diffraction limit: near- and far-field optical encryption," Nanophotonics **10**, 697–704 (2021). https://doi.org/10.1515/nanoph-2020-0352.
23. T. O'Connor, A. Doblas, and B. Javidi "Structured illumination in compact and field-portable 3D-printed shearing digital holographic microscopy for resolution enhancement," Opt. Lett. **44**, 2326–2329 (2019). https://doi.org/10.1364/OL.44.002326.
24. F. Borrelli, J. Behal, A. Cohen, *et al.*, "AI-aided holographic flow cytometry for label-free identification of ovarian cancer cells in the presence of unbalanced datasets," APL Bioeng. **7**, 026110 (2023). https://doi.org/10.1063/5.0153413.
25. R. Guo, S. K. Mirsky, I. Barnea, *et al.*, "Quantitative phase imaging by wide-field interferometry with variable shearing distance uncoupled from the off-axis angle," Front. Phys. **28**, 5617–5628 (2020). https://doi.org/10.1364/OE.385437.
26. J. Behal, "Quantitative phase imaging in common-path cross-referenced holographic microscopy using double-exposure method," Sci. Reports **9**, 9801 (2019). https://doi.org/10.1038/s41598-019-46348-9.
27. R. Guo, I. Barnea, and N. T. Shaked, "Low-Coherence Shearing Interferometry With Constant Off-Axis Angle," Front. Phys. **8**, 611679 (2021). https://doi.org/10.3389/fphy.2020.611679.
28. J. W. Goodman, "Statistical Optics, 2nd Edition," John Wiley & Sons, New York (2015). ISBN: 978-1-119-00945-0. https://www.wiley.com/en-ca/Statistical+Optics%2C+2nd+Edition-p-9781119009450.
29. M. R. Vogt, H. Hahn, H. Holst, *et al.*, "Measurement of the Optical Constants of Soda-Lime Glasses in Dependence of Iron Content and Modeling of Iron-Related Power Losses in Crystalline Si Solar Cell Modules," IEEE J. Photovoltaics **6**, 111–118 (2016). https://doi.org/10.1109/JPHOTOV.2015.2498043.
30. P. Memmolo, L. Miccio, M. Paturzo, *et al.*, "Recent advances in holographic 3D particle tracking," Adv. Opt. Photon. **7**, 713–755 (2015). https://doi.org/10.1364/AOP.7.000713.